# Anomalous Hall voltage rectification and quantized spin-wave excitation induced by the simultaneous dc- and rf-current application in $Ni_{81}Fe_{19}$ wire

A. Yamaguchi[1, 2], K. Motoi[1], A. Hirohata[3] and H. Miyajima[1]

[1)] Department of Physics, Keio University, Hiyoshi, Yokohama 223-8522, Japan

[2)] PRESTO, JST, Honcho, Kawaguchi, Saitama 332-0012, Japan

[3)] Department of Electronics, University of York, Heslington, York YO10 5DD, England

* Corresponding authors' E-mail: yamaguch@phys.keio.ac.jp (A. Y.), kmotoi@phys.keio.ac.jp (K.M.) and miyajima@phys.keio.ac.jp (H. M.)

**[Abstract]**

An anomalous Hall effect and rectification of a Hall voltage are observed by applying a radio-frequency (rf) current through a single-layered ferromagnetic wire located on a coplanar waveguide. The components of the magnetization precession, both in and perpendicular to the plane, can be detected via the Hall voltage rectification of the rf current by incorporating an additional direct (dc) current. In this paper, we propose a phenomenological model, which describes the time-dependent anisotropic magnetoresistance and time-dependent planer Hall effect. The nonlinearity of the spin dynamics accompanied by spin-waves as functions of rf and dc currents is also studied, as well as those of the magnitude and orientation of the external magnetic field.

# I. INTRODUCTION

The understanding of spin dynamics in artificial nano-magnets is vital, not only for fundamental magnetism but also technological applications. One of the distinctive characteristics of the spin dynamics within a high-frequency region is spin-wave (SW) excitation, in which an extended ringing of the magnetization produces a high-frequency resonance characteristic within a device[1, 2]. For example, the progress in magnetic recording spawns the development of smaller patterned media and a faster read-write time. Consequently, within such nano-scale ferromagnetic devices, both exchange and dipole energies contribute to SW excitation, both of which are strongly dependent on the system geometry[1, 2]. Accordingly, the SW resonance related to magnetization dynamics in a confined geometry has been investigated by using Brillouin light scattering (BLS)[1 - 3], ferromagnetic resonance (FMR)[1, 4], the time-resolved magneto-optical Kerr effect[5, 6], and the rectification effect[7 - 12].

When the spin-polarized current flows through a ferromagnetic conductor, the spin angular momentum of the conduction electrons transfers to the magnetic moment with the assistance of the spin-transfer effect, and, consequently, the local magnetization precesses. Highly sensitive electrical detection of the magnetization dynamics is achieved by using such spin-transfer torque induced by a spin-polarized alternating current (ac current)[8, 9]. One of the most interesiting discoveries of the rectification of the radio-frequency (rf) electromagnetic field is the direct-current (dc) voltage spectrum measured with respect to the magnetization dynamics and the magnitude of the spin-transfer torque[8, 9]. The strength and direction of the spin-transfer torque are estimated by using the rectification effect in a magnetic tunneling junction. These also lead to a detailed understanding of the spin dynamics induced by the interactions between conduction electrons and magnetic moments.

About fifty years ago, Juretchke[7] discovered the rectification effect and the Hall effect in thin magnetic film in its FMR state, which is a consequence of the magnetoresistance oscillation induced by a microwave electromagnetic driving field. The effect offers highly sensitive and simple detection of magnetization dynamics. In this paper, we extend and develop the method used to investigate the magnetization dynamics in nano- or micron-scale confined magnets, showing that the Hall voltage rectification is directly coupled with a dc current.

The propagation of electromagnetic waves through a ferromagnetic conductor raises some galvanomagnetic effects, reflecting the interactions between electrical current and magnetization as described in the magnetoresistance and extraordinary Hall effects. The galvanomagnetic effects remarkably emerge in the vicinity of the FMR frequency and can be measured as electrical signals. We have conducted FMR studies on a single-layered $Ni_{81}Fe_{19}$ (Py) ferromagnetic wire under the simultaneous application of both dc and rf currents. The next question is what the spin-torque excites when only the dc spin-polarized current is applied to the SW excited state with an inhomogeneous spin distribution. For example, both the adiabatic spin-transfer and non-adiabatic torques have been experimentally confirmed as displacing the domain wall (DW)[16-18], to excite the quantized SW[19, 20] and induce the SW Doppler shift[21]. Further investigation into such responses offers a deeper understanding of the spin dynamics correlated with the spin transport in inhomogeneous magnetization distributions[22 - 25]. Even so, the ground state under the presence of the dc spin current, inducing the instability of a uniformly magnetized state[21, 26, 27], remains to be identified.

The rectification effect allows us to perform highly sensitive detection of the small friction of the spin dynamics within a nano- or micron-scale confined magnet. Therefore, we propose a phenomenological model for the magnetoresistance response induced by constant-wave (CW) microwave excitation and dc current, and also present a consistent view of dc voltage generation in ferromagnetic wire. This model is applied to the spin dynamics induced by both the dc and rf currents, which are measured as dc spectra in a well-resolved frequency-domain. A perpendicular standing spin wave (PSSW) in addition to a quantized in-plane SW is detected as the rectification of the planer Hall effect (PHE). The generation of a Hall voltage signal is expected to be expressed as either $V_{\mathrm{dc+rf}}^{\mathrm{Hall}} = V_{\mathrm{dc}}^{\mathrm{Hall}} + V_{\mathrm{rf}}^{\mathrm{Hall}}$ or $V_{\mathrm{dc+rf}}^{\mathrm{Hall}} = V_{\mathrm{dc}}^{\mathrm{Hall}} + V_{\mathrm{rf}}^{\mathrm{Hall}} + \delta V_{\mathrm{dc \bullet rf}}^{\mathrm{Hall}}$, where $V_{\mathrm{dc}}^{\mathrm{Hall}}$, $V_{\mathrm{rf}}^{\mathrm{Hall}}$ and $\delta V_{\mathrm{dc \bullet rf}}^{\mathrm{Hall}}$ represent the dc Hall voltage, the rf Hall voltage and the mixing term which corresponds to the additional Hall voltage induced by coupling with the dc and rf currents.

In this paper, we present an experimental investigation into the Hall voltage rectification effect due to the magnetization dynamics in a single-layered ferromagnetic micron-scale Py wire. The present experimental setup is presented in Section II, while Section III provides the analytical model concerning the Hall voltage based on the magnetoresistance oscillations by the combined application of both the dc and rf currents. The experimental results of the Hall voltage measured in the ferromagnetic wire are described in Section IV. We focus on their characteristic dependences on both the dc current and the magnetic field orientation, both of which are assessed by the present model, while in Section V, the conclusions are summarized.

## II. EXPERIMENTAL SETUP

A 65 nm thick Py wire, 150 μm long and 5 μm wide, was fabricated on an MgO substrate via a combination of ultrahigh vacuum deposition, electron beam lithography and the lift-off method[10]. Figure 1(a) shows an optical micrograph of the wire together with an electric measurement circuit. The wire was placed on the center conductive strip line within the coplanar waveguide (CPW) structure. A sinusoidal CW rf current with a current density of $1.5\times10^{10}$ A/m$^2$ was subsequently injected into the wire by a signal generator within the frequency range 10 MHz and 15 GHz. Simultaneously, a dc current was applied to the above rf current via the bias-tee, which separates the dc- and rf-components of the current. The external magnetic field $\boldsymbol{H}_{\mathrm{ext}}$ was also applied to the substrate plane as a function of angle $\phi$ from the major axis of the wire. The magnetization precession in the vicinity of the FMR state induced anisotropic magnetoresistance (AMR) oscillation[7 - 12] and also generated the dc voltage $V_{\mathrm{AMR}}$. The experiment was performed at room temperature with the slowly sweeping frequency of the rf current flowing along the major axis of the wire. The Hall voltage spectra $V_{\mathrm{Hall}}$, induced across the minor axis of the wire, were also simultaneously measured.

# III. ANALYTICAL MODEL

The electrical conduction in a ferromagnet generally depends on the direction of the magnetization, and the phenomenological relationship between the voltage $\boldsymbol{E}$ and the electrical current density $\boldsymbol{j}$ is written as:[7]

$$\boldsymbol{E} = \rho_{\perp}\boldsymbol{j} + \boldsymbol{m}\left(\boldsymbol{j}\cdot\boldsymbol{m}\right)\cdot\left(\rho_{\parallel} - \rho_{\perp}\right) + \rho_{\mathrm{H}}\boldsymbol{m}\times\boldsymbol{j}, \tag{1}$$

where $\boldsymbol{m}$ is the unit vector along the local magnetization, $\rho_{\perp}$ and $\rho_{\parallel}$ are the resistivities perpendicular and parallel to $\boldsymbol{j}$, respectively, and $\rho_{\mathrm{H}}$ is the extraordinary Hall resistivity. Juretschke[7] has introduced an oscillating component of the magnetization $\boldsymbol{m} = \boldsymbol{m}_0 + \delta\boldsymbol{m}\left(t\right)$ into Eq. (1), and pointed out that a dc voltage is generated in the magnetization precession induced by the rf electric field.

The frequency spectra of the SW excitations in a ferromagnetic wire are evaluated by using the analytical model proposed by Guslienko *et al*[28]. Using their model, in order to focus on the essence of the phenomena, we introduce a simplified phenomenological analysis based on the macro-spin model, which corresponds to the SW mode with the lowest index. As shown in the coordinate system in Fig. 1(b), when the magnetization unit vector at the origin is $\boldsymbol{m} = \left(\sin\theta\cos\phi, \sin\theta\sin\phi, \cos\theta\right)$ directing along the effective magnetic field orientation and the electrical current flows along the longitudinal axis of the wire as $\boldsymbol{j} = \left(j(t), 0, 0\right)$, the electric field $\boldsymbol{E}$ is given by:

$$\boldsymbol{E}(t) = \begin{pmatrix} E_{\mathrm{x}}(t) \\ E_{\mathrm{y}}(t) \\ E_{\mathrm{z}}(t) \end{pmatrix} = j(t) \begin{pmatrix} \rho_{\perp} + \Delta\rho\sin^2\theta\cos^2\phi \\ \Delta\rho\sin^2\theta\cos\phi\sin\phi + \rho_{\mathrm{H}}\cos\theta \\ \Delta\rho\sin\theta\cos\theta\cos\phi - \rho_{\mathrm{H}}\sin\theta\sin\phi \end{pmatrix}, \quad (2)$$

where $\Delta\rho = \rho_{\parallel} - \rho_{\perp}$. Here, the current has dc and rf components: $j(t) = j_{\mathrm{dc}} + j_{\mathrm{rf}}\,\mathrm{Re}\left(e^{-i\omega t}\right) = j_{\mathrm{dc}} + j_{\mathrm{rf}}\cos\omega t$, where $j_{\mathrm{dc}}$ and $j_{\mathrm{rf}}$ denote the dc and rf current densities, respectively, and $\omega$ is the angular frequency of the rf excitation current. To simplify the electric field described in Eq. (2), the magnetization is assumed to point along the effective magnetic field in the substrate plane due to the strong magnetic shape anisotropy and the external static magnetic field. Namely, $\theta = 90^{\circ}$. When the rf current in addition to the dc current is injected into the $Ni_{81}Fe_{19}$ wire, the magnetization precession is induced by its driving torque, including the spin-transfer effect and the non-uniform magnetic field[13, 14, 21-24]. Consequently, the time-dependent electric fields, $E_{\mathrm{x}}$, $E_{\mathrm{y}}$ and $E_{z}$ along the major axis (*x*-axis), the minor-axis (*y*-axis) and the vertical-axis (*z*-axis) of the wire due to the precession are respectively given by

$$E_x(t) = j(t)\left\{\rho_\perp + \Delta\rho\cos^2\left(\phi + \delta(t)\right)\right\}, \tag{3}$$

$$E_y(t) = j(t)\Delta\rho\cos\left(\phi + \delta(t)\right)\sin\left(\phi + \delta(t)\right) \tag{4}$$

and

$$E_z(t) = -j(t)\rho_{\mathrm{H}}\sin\left(\phi + \delta(t)\right), \tag{5}$$

where $\delta(t)$ is the magnetization precessional angle around the effective field and is generally defined so as to involve both in and out-of-plane excursions of the magnetization, which differ significantly due to the demagnetizing-field-induced ellipticity of the magnetization precession. The individual contributions are discussed later in this Section. Equations (3)-(5) indicate the AMR effect, the PHE and the anisotropic Hall effect, respectively. By expanding Eqs. (3)-(5) using the addition formula of a trigonometric function, we can derive an average time-dependent electric field to the second order of the dynamic magnetization precessional angle $\delta(t)$ as follows:

$$E_{\mathrm{x}}(t) \approx j(t)\left[\rho_\perp + \Delta\rho\left(\cos^2\phi - \frac{1}{2}\sin 2\phi\sin 2\delta(t) - \cos 2\phi\sin^2\delta(t)\right)\right], \tag{6}$$

$$E_{\mathrm{y}}(t) \approx \frac{1}{2}j(t)\Delta\rho\left[\sin 2\phi + \cos 2\phi\sin 2\delta(t) - 2\sin 2\phi\sin^2\delta(t)\right] \tag{7}$$

and

$$E_z(t) \approx -j(t)\rho_{\mathrm{H}}\left[\sin\phi + \cos\phi\sin\delta(t) - \frac{1}{2}\sin\phi\sin^2\delta(t)\right]. \tag{8}$$

These relations clearly illustrate that the time variation of the electric fields $E_{\mathrm{x}}(t)$, $E_{\mathrm{y}}(t)$ and $E_z(t)$ is induced by the magnetization precession and the injection of the time-dependent microwave field.

To understand the phenomena qualitatively, we treat the magnetization dynamics in a simple state, where the magnetization only precesses around the effective magnetic field direction as mentioned above. The previous models are based on the assumption that the SW excitation can be described by the smooth undulation and small amplitude of the magnetization. In the present model, the Landau-Lifshitz-Gilbert (LLG) equation is considered, including the spin-transfer torque and the rf magnetic field, which consists of both an inhomogeneous electromagnetic field and a dynamic demagnetizing field due to the SW excitation. The LLG equation to describe the magnetization dynamics is therefore written by[21 - 25]

$$\frac{\partial \boldsymbol{m}}{\partial \mathrm{t}} = -\gamma_0 \boldsymbol{m} \times \left( \boldsymbol{H}_{\mathrm{eff}} + \boldsymbol{h}_{\mathrm{rf}} \right) + \alpha \boldsymbol{m} \times \frac{\partial \boldsymbol{m}}{\partial \mathrm{t}} - \left( \boldsymbol{u} \cdot \nabla \right) \boldsymbol{m} + \beta \boldsymbol{m} \times \left[ \left( \boldsymbol{u} \cdot \nabla \right) \boldsymbol{m} \right], \tag{9}$$

where $\boldsymbol{m}(t)$ denotes the unit vector along the local magnetization ($\boldsymbol{m} = \boldsymbol{M}/M_{\mathrm{S}}$, $|\boldsymbol{m}| = 1$ and $M_{\mathrm{S}}$: saturation magnetization), and $\gamma_0$, $\boldsymbol{H}_{\mathrm{eff}}$, $\boldsymbol{h}_{\mathrm{rf}}$ and $\alpha$ represent the gyromagnetic ratio, the effective magnetic field including exchange and demagnetizing fields, the rf field produced by the rf current flowing through the middle strip of the CPW and the Gilbert damping constant, respectively. Furthermore, $\boldsymbol{u}$ is given by using the current density $\boldsymbol{j}$ and the spin polarization of the current $P$ as follows[21 - 25]:

$$\boldsymbol{u} = \boldsymbol{j} \frac{P \mu_{\mathrm{B}}}{e M_{\mathrm{S}}}. \tag{10}$$

Here, the previous approaches have been developed based on the LLG equation, while the present treatment includes one additional important aspect; the spin-transfer torque in the SW excitation state gives the precession of the precessional axis of the magnetization itself[29]. Therefore, we introduce the angular vector $\mathbf{\Omega}$ and replace the time derivative of $\boldsymbol{m}(t)$ with the following:

$$\frac{\partial \boldsymbol{m}(t)}{\partial t} \Rightarrow \frac{\partial \boldsymbol{m}(t)}{\partial t} + \mathbf{\Omega} \times \boldsymbol{m}(t). \tag{11}$$

Consequently, Eq. (9) is rewritten as

$$\frac{\partial \boldsymbol{m}}{\partial t} + \mathbf{\Omega} \times \boldsymbol{m} = -\gamma_0 \boldsymbol{m} \times \left(\boldsymbol{H}_{\text{eff}} + \boldsymbol{h}_{\text{rf}}\right) + \alpha \boldsymbol{m} \times \left(\frac{\partial \boldsymbol{m}}{\partial t} + \mathbf{\Omega} \times \boldsymbol{m}\right) - \left(\boldsymbol{u} \cdot \nabla\right)\boldsymbol{m} + \beta \boldsymbol{m} \times \left[\left(\boldsymbol{u} \cdot \nabla\right)\boldsymbol{m}\right]. \tag{12}$$

In the right-hand side of Eqs. (9) and (12), the second term represents the damping effect, and the third and fourth terms correspond to the spin-transfer torque and spin-flip of the conduction electrons, respectively[21-25]. From Eqs. (9) and (12), the magnetization gradient along the electric field $-\left(\boldsymbol{u} \cdot \nabla\right)\boldsymbol{m} + \beta \boldsymbol{m} \times \left[\left(\boldsymbol{u} \cdot \nabla\right)\boldsymbol{m}\right]$ is obtained when the spin-transfer torque and spin-flip term dominate the magnetization dynamics. The vector equation (12) can be linearized and projected onto the two normal vectors ***b*** and ***c*** as shown in Fig. 1(c). Here, Eq. (12) for the case of $\theta = 90^{\circ}$ is considered since the magnetization nearly lies in the plane and aligns along the major axis (*x*-axis) due to the sufficiently strong magnetic shape anisotropy. In defining the present coordinate system, ***b*** is in the *x-y* plane and ***c*** is perpendicular to the *x-y* plane. Now, $\boldsymbol{m} = \boldsymbol{m}_0 + \delta \boldsymbol{m}$ with $|\boldsymbol{m}| = 1$, where $\boldsymbol{m}_0 = (m_0, 0, 0)$ corresponds to the equilibrium direction of the magnetization along the effective magnetic field (*a*-axis), and the small deviation $\delta \boldsymbol{m}(t)$ is $\left(0, m_b(t), m_c(t)\right) \approx \left(0, m_b \mathrm{e}^{i\omega t}, m_c \mathrm{e}^{i\omega t}\right)$. The angular vector $\boldsymbol{\Omega}$ is given by $\left(0, 0, \Omega_0\right)$ since the magnetization precesses in the plane, while the spin current $\boldsymbol{u}$ is considered to be spatially uniform along the longitudinal axis of the wire (*x*-axis). With the axes defined in Fig. 1, $\boldsymbol{u}$ is given by $\boldsymbol{u} = \left(u(t), 0, 0\right) = \left(u_{\mathrm{dc}} + u_{\mathrm{rf}} \cos \omega t, 0, 0\right)$ in the (*x*, *y*, *z*) coordinate system. The adiabatic and non-adiabatic spin-transfer torque terms are therefore obtained in the (*a*, *b*, *c*) coordinate system[29] to the first order of the deviations $\delta \boldsymbol{m}(t)$:

$$-(\boldsymbol{u} \cdot \nabla)\boldsymbol{m} + \beta \boldsymbol{m} \times \left[(\boldsymbol{u} \cdot \nabla)\boldsymbol{m}\right] = u(t) \frac{\partial \phi}{\partial x} \begin{pmatrix} 0 \\ -1 \\ \beta \end{pmatrix}. \tag{13}$$

Next, the dynamic and static magnetic field terms and the damping term are calculated. They are described as the first and second terms in the right-hand side of Eq. (12), respectively. The external magnetic field is directed at the angle $\phi_0$ from the $+x$ coordinate axis. Subsequently, we redefine a new $(a, b, c)$ coordinate system, where the $+a$ direction corresponds to the equilibrium direction of $\boldsymbol{m}_0$ along the effective magnetic field $\boldsymbol{H}_{\mathrm{eff}} = \boldsymbol{H}_{\mathrm{ext}} + \boldsymbol{H}_{\mathrm{A}}$ ($\boldsymbol{H}_{\mathrm{ext}}$: an external field and $\boldsymbol{H}_{\mathrm{A}}$: a shape anisotropy field).

The magnetization precession around $\boldsymbol{H}_{\mathrm{eff}}$ results in a small time-dependent component of the magnetization perpendicular to $\boldsymbol{m}_0$, which inclines at the angle $\phi$ from the $+x$ axis. The magnetic fields, $\boldsymbol{H}_{\mathrm{eff}}$, $\boldsymbol{H}_{\mathrm{ext}}$ and $\boldsymbol{H}_{\mathrm{A}}$, in the $(a, b, c)$ coordinate system satisfy the following relationship[30]:

$$\boldsymbol{H}_{\text{eff}} = \boldsymbol{H}_{\text{ext}} + \boldsymbol{H}_{\text{A}} . \tag{14}$$

Here,

$$\boldsymbol{H}_{\text{ext}} = \left( H_{\text{ext}} \cos\left(\phi_0 - \phi\right),\, H_{\text{ext}} \sin\left(\phi_0 - \phi\right),\, 0 \right), \tag{15}$$

and

$$\boldsymbol{H}_{\text{A}} = -M_{\text{S}} \tilde{\boldsymbol{N}} \cdot \boldsymbol{m} , \tag{16}$$

where $\tilde{\boldsymbol{N}}$ is the demagnetizing-factor tensor in the $(a,b,c)$ coordinate system, which is given by[27]

$$\tilde{\boldsymbol{N}} = \begin{pmatrix} N_x \cos^2\phi + N_y \sin^2\phi & \frac{1}{2}\left(N_x - N_y\right)\sin 2\phi & 0 \\ \frac{1}{2}\left(N_x - N_y\right)\sin 2\phi & N_x \sin^2\phi + N_y \cos^2\phi & 0 \\ 0 & 0 & N_z \end{pmatrix}, \tag{17}$$

where $N_x, N_y$ and $N_z$ are the demagnetizing factors in the $(x,y,z)$ system. It should be noted that Equation (17) satisfies the Schlomann sum rule[31]: $\mathrm{Tr}\tilde{\boldsymbol{N}} = 1$ . Therefore, the contribution of the static magnetic field is obtained as follows:

$$-\gamma_0 \boldsymbol{m} \times \left( \boldsymbol{H}_{\text{ext}} + \boldsymbol{H}_{\text{A}} \right) \approx -\gamma_0 \begin{pmatrix} -m_c e^{i\omega t} H'_{\text{eff}} \\ m_c e^{i\omega t} H'_c \\ H'_{\text{eff}} - m_b e^{i\omega t} H'_b \end{pmatrix}, \tag{18}$$

where

$$H'_b = H_{\text{ext}} \cos\left(\phi_0 - \phi\right) + M_{\text{S}} \left(N_y - N_x\right) \cos 2\phi , \tag{19}$$

$$H'_c = H_{\text{ext}} \cos\left(\phi_0 - \phi\right) + M_{\text{S}} \left[ N_z - \left( N_x \cos^2\phi + N_y \sin^2\phi \right) \right] \tag{20}$$

and

$$H'_{\text{eff}} = H_{\text{ext}} \sin\left(\phi_0 - \phi\right) - \frac{1}{2} M_{\text{S}} \left(N_x - N_y\right) \sin 2\phi . \tag{21}$$

Here, the rf field $\boldsymbol{h}_{\text{rf}}$ is given by

$$\boldsymbol{h}_{\text{rf}} = \mathrm{e}^{i\omega t} \left(h_{\text{in}} \sin\phi, h_{\text{in}} \cos\phi, h_{\text{out}}\right), \tag{22}$$

where $h_{\text{in}}$ and $h_{\text{out}}$ denote the in-plane and out-of-plane fields, respectively. The rf field is composed of not only the magnetic field component of the rf electromagnetic wave but also the Oersted field produced by the rf current flowing through the wire and CPW electrode[30]. Subsequently, the dynamic magnetic torque and damping terms are respectively obtained as follows:

$$-\gamma_0 \boldsymbol{m} \times \boldsymbol{h}_{\text{rf}} = e^{i\omega t} \begin{pmatrix} 0 \\ \gamma_0 h_{\text{out}} \\ -\gamma_0 h_{\text{in}} \cos\phi \end{pmatrix} \tag{23}$$

and

$$\alpha \boldsymbol{m} \times \left( \frac{\partial \boldsymbol{m}}{\partial t} + \boldsymbol{\Omega} \times \boldsymbol{m} \right) = e^{i\omega t} \begin{pmatrix} -\alpha \Omega_0 m_c \\ -i\omega \alpha m_c \\ i\omega \alpha m_b + \Omega_0 \alpha e^{-i\omega t} \end{pmatrix}. \tag{24}$$

By substituting Eqs. (13), (18), (23) and (24) into Eq. (12) and neglecting the minor quadratic terms, we obtain

$$\frac{m_b}{m_c} = \left( \alpha - \frac{\gamma_0 H'_{\text{eff}}}{\Omega_0} \right), \tag{25}$$

and

$$\begin{pmatrix} i\omega & [\gamma_0 H'_c + i\omega\alpha] \\ -[\gamma_0 H'_b + i\omega\alpha] & i\omega \end{pmatrix} \begin{pmatrix} m_b \\ m_c \end{pmatrix} = e^{-i\omega t} \begin{pmatrix} -\Omega_0 - u_{\mathrm{dc}} \dfrac{\partial \phi}{\partial x} \\ \alpha\Omega_0 + \beta u_{\mathrm{dc}} \dfrac{\partial \phi}{\partial x} \end{pmatrix} + \gamma_0 \begin{pmatrix} h_{\mathrm{out}} \\ -h_{\mathrm{in}} \cos\phi \end{pmatrix} + u_{\mathrm{rf}} \frac{\partial \phi}{\partial x} \begin{pmatrix} -1 \\ \beta \end{pmatrix}. \tag{26}$$

The component $m_b$ is obtained by solving Eq. (26);

$$m_b = \frac{1}{\det|\chi|} \left\{ e^{-i\omega t} \left[ i\omega A_1 - \left(\gamma_0 H'_c + i\omega\alpha\right) B_1 \right] + \left[ i\omega A_2 - \left(\gamma_0 H'_c + i\omega\alpha\right) B_2 \right] \right\}, \tag{27}$$

where the driving terms $A_1, B_1, A_2, B_2$ and $\det|\chi|$ are respectively given by:

$$A_1 = -\Omega_0 - u_{\mathrm{dc}} \frac{\partial \phi}{\partial x}, \tag{28}$$

$$B_1 = \alpha\Omega_0 + \beta u_{\mathrm{dc}} \frac{\partial \phi}{\partial x}, \tag{29}$$

$$A_2 = \gamma_0 h_{\mathrm{out}} - u_{\mathrm{rf}} \frac{\partial \phi}{\partial x}, \tag{30}$$

$$B_2 = -\gamma_0 h_{\mathrm{in}} \cos\phi + \beta u_{\mathrm{rf}} \frac{\partial \phi}{\partial x}, \tag{31}$$

$$\det|\chi| = -\omega^2 + \left[\gamma_0 H'_c + i\omega\alpha\right] \cdot \left[\gamma_0 H'_b + i\omega\alpha\right], \tag{32}$$

and the FMR frequency $\omega_k$ and full width at half maximum $\Delta\alpha$ are written by:

$$\omega_k^2 = \gamma_0^2 H'_c H'_b \tag{33}$$

and

$$\Delta\alpha = \gamma_0 \left(H'_b + H'_c\right)\alpha. \tag{34}$$

Taking the complex conjugates into account, the $m_b$ in the FMR state is derived as

$$m_b(t) = D_0 + D_1 \cos\omega_k t + D_2 \sin\omega_k t, \tag{35}$$

where the coefficients $D_0, D_1$ and $D_2$ are given by the following relations, respectively:

$$D_0 = \frac{A_1 - B_1}{\alpha\Delta} = -\frac{1}{\alpha\Delta} \left[ \Omega_0 \left(1+\alpha\right) + \left(1+\beta\right) u_{\mathrm{dc}} \frac{\partial \phi}{\partial x} \right], \tag{36}$$

$$D_1 = \frac{A_2 - \alpha B_2}{\alpha\Delta} = \frac{1}{\alpha\Delta} \left[ \gamma_0 \left(h_{\mathrm{out}} + \alpha h_{\mathrm{in}} \cos\phi\right) - \left(1+\alpha\beta\right) u_{\mathrm{rf}} \frac{\partial \phi}{\partial x} \right], \tag{37}$$

and

$$D_2 = -\frac{\gamma_0 H'_c}{\omega_k \alpha \Delta} B_2 = -\frac{\gamma_0 H'_c}{\omega_k \alpha \Delta}\left(-\gamma_0 h_{\text{in}} \cos\phi + \beta u_{\text{rf}} \frac{\partial \phi}{\partial x}\right). \tag{38}$$

The first term in the right-hand side of Eq. (35) describes the magnetization precession due to the dc spin transfer effect, while the second and last terms show the in-phase and out-of-phase driving torque, respectively. The space derivative term $\partial\phi/\partial x$ is introduced as a phenomenological parameter of the spin-transfer terms. The individual components of the effective field, including the demagnetizing factor of the wire, $H'_b, H'_c$ and $H'_{\text{eff}}$, are derived in Eqs. (19)-(21), respectively, which correspond to the terms described in our previous paper[29]. The adequately small precessional angle $\delta(t)$ is given by $\boldsymbol{m}$, and $\delta\boldsymbol{m}(t)$ satisfies $\sin\delta(t) \approx \left\langle \sqrt{m_b^2 + m_c^2} \right\rangle / |\boldsymbol{m}|$. The out-of-plane component $m_c$ generates the dynamic demagnetizing field and exerts torque proportional to $\boldsymbol{m} \times \delta\boldsymbol{m}$ onto the magnetization, rotating $\boldsymbol{m}$ by the angle $\delta(t)$ in the plane.

Considering the relationship between $m_b$ and $m_c$ as described in Eq. (25), we obtain the following relationship:

$$m_c \approx -\eta m_b \ , \tag{39}$$

where the coefficient $|\eta|$ is estimated at 0.04 because the damping constant $\alpha$ of $Ni_{81}Fe_{19}$ is typically 0.01. This provides $\gamma_0 H'_{\text{eff}} \approx 350$ MHz (when $\phi \approx \phi_0, N_x \approx 0, N_y \approx 0.05, \text{and}\, N_z \approx 0.95$) as estimated by Eq. (21). $\Omega_0$ is also estimated to be $\omega_k \approx 6.3$ GHz, which corresponds to the FMR frequency given by Eq. (33)[1, 2, 28, 30,32]. These results indicate that the magnetization precesses around the effective field direction with a highly elliptical orbit in plane. According to Eq. (39), $\delta(t)$ satisfies

$$\sin\delta(t) \approx \frac{m_b\sqrt{1+\eta^2}}{|\boldsymbol{m}|} = m_b\sqrt{1+\eta^2} \text{ and } \sin 2\delta(t) \approx 2m_b\sqrt{1+\eta^2} . \tag{40}$$

Following the substitution of Eqs. (35) and (40) into Eqs. (6)-(8), the time variation of the individual electric fields $E_x(t), E_y(t)$ and $E_z(t)$ is calculated. Here, in the FMR state ($\omega = \omega_k$), the experimentally measured voltages are given by the time average of $\langle E_x(t)\rangle, \langle E_y(t)\rangle$ and $\langle E_z(t)\rangle$. As $\langle\cos^2\omega t\rangle = \langle\sin^2\omega t\rangle = 1/2$, the time independent voltages are given by

$$\begin{aligned}\langle E_{\mathrm{x}}(t)\rangle \approx\, & j_{\mathrm{dc}}\left[\rho_{\perp} + \Delta\rho\cos^2\phi - \sqrt{1+\eta^2}D_0\Delta\rho\sin 2\phi - \left(1+\eta^2\right)\left(D_0^2 + \frac{D_1^2+D_2^2}{2}\right)\Delta\rho\cos 2\phi\right] \\ & - j_{\mathrm{rf}}\left[\frac{\Delta\rho}{2}\sqrt{1+\eta^2}D_1\sin 2\phi - \left(1+\eta^2\right)D_0D_1\cos 2\phi\right]\end{aligned}, \tag{41}$$

$$\begin{aligned}\langle E_{\mathrm{y}}(t)\rangle \approx\, & \frac{1}{2}j_{\mathrm{dc}}\Delta\rho\left[\sin 2\phi + 2\sqrt{1+\eta^2}D_0\cos 2\phi - 2\left(1+\eta^2\right)\left(D_0^2 + \frac{D_1^2+D_2^2}{2}\right)\sin 2\phi\right] \\ & + \frac{1}{2}j_{\mathrm{rf}}\Delta\rho\left[\sqrt{1+\eta^2}D_1\cos 2\phi - 2\left(1+\eta^2\right)D_0D_1\sin 2\phi\right]\end{aligned} \tag{42}$$

and

$$\begin{aligned}\langle E_z(t)\rangle \approx\, & j_{\mathrm{dc}}\rho_{\mathrm{H}}\left[\sin\phi + \sqrt{1+\eta^2}D_0\cos\phi - \frac{1}{2}\left(1+\eta^2\right)\left(D_0^2 + \frac{D_1^2+D_2^2}{2}\right)\sin\phi\right] \\ & + \frac{1}{2}j_{\mathrm{rf}}\rho_{\mathrm{H}}\left[\sqrt{1+\eta^2}D_1\cos\phi - \left(1+\eta^2\right)D_0D_1\sin\phi\right]\end{aligned}, \tag{43}$$

By substituting Eqs. (36)-(38) into Eqs. (41) and (42), the field angle $\phi_0$ dependence of the individual variations in the rectified signals is obtained. Here, if there is a uniform spin structure in the wire and $\partial\phi/\partial x$, its variation with $x$ is negligible, and the spin torque does not play an important role. Then, the spin dynamics under the spin torque does not occur along the $x$ direction. In other words, $\Omega_0 = 0$, which means $D_0 \approx 0$. In previous studies, therefore, Eqs. (41)-(43) are approximated to the first order of $D_1$, and only the rectifying voltage in combination with the rf current and magnetization dynamics are under consideration.

Here, the second order of $D_n\left(n = 0,1,2\right)$ should be reserved since the second terms of the right-hand side in Eqs. (41)-(43) provide interplaying contributions between the dc current $j_{\mathrm{dc}}$ and the rf current $j_{\mathrm{rf}}$. The terms cannot be negligible when both the dc and rf currents are simultaneously applied. Although making only minute contributions to the dc voltage spectra, they still play an important role in distinguishing individual contributions to the various driving torques, as shown in Section IV D.

For example, the magnetization dynamics induced by the in-plane rf field $h_{\mathrm{in}}$ due to the microwave introduction (*i.e.* $j_{\mathrm{dc}} = 0$ and $j_{\mathrm{rf}} \neq 0$) gives the amplitude of the dc voltages $V_{\mathrm{AMR}}\left(\omega_k\right)$ and $V_{\mathrm{Hall}}\left(\omega_k\right)$ by Eqs. (41) and (42) in the FMR state, corresponding to the AMR voltage generated along the major (*x*) axis and the Hall voltage along the minor (*y*) axis as follows:

$$\Delta V_{\mathrm{AMR}}\left(\omega_k, I_{\mathrm{dc}} = 0\,\mathrm{mA}\right) \propto \sin 2\phi \cdot \cos\phi\,, \tag{44}$$

$$\Delta V_{\mathrm{Hall}}\left(\omega_k, I_{\mathrm{dc}} = 0\,\mathrm{mA}\right) \propto \cos 2\phi \cdot \cos\phi\,. \tag{45}$$

# IV. EXPERIMENTAL RESULTS AND DISCUSSION

### A. Simultaneous measurement of the voltages generated along the major and minor axes

Firstly, the rectification spectra generated along the *x*- and *y*-axes are simultaneously measured so as to confirm whether anomalous behaviors are produced by the microwave distribution.

An example of representative spectra at $H_{\mathrm{ext}} = 500$ Oe and $\phi = 45^{\circ}$ is shown in Fig. 2(a), where the voltage amplitude in the FMR state is defined as a peak-to-peak value as shown in the figure. The angle $\phi$ dependence of $\Delta V_{\mathrm{AMR}}$ and $\Delta V_{\mathrm{Hall}}$ is shown in Figs. 2(b) and (c) together with the curve fitted to Eqs. (44) and (45). As is seen in the figure, the $\phi$ dependence in the case of low field ($H_{\mathrm{ext}} = 100\,\mathrm{Oe}$) does not seem to agree with the calculation, since the uniform precession is difficult to realize when the magnetization is not directed along the low external-field. Bailleul *et al.*[33] have reported that the non-zero *y* component of the exchange field proportional to $\mathrm{d}^2 M_{\mathrm{y}} / \mathrm{d}x^2$ is attributed to a non-collinear magnetization alignment in the vicinity of the edge. This leads to the departure of the angular dependences of $\Delta V_{\mathrm{AMR}}$ and $\Delta V_{\mathrm{Hall}}$ from the analytical fitting curve. Conversely, where $H_{\mathrm{ext}} = 500\,\mathrm{Oe}$ exceeding the shape anisotropy field, the curve fitting to Eqs. (44) and (45) virtually correspond to the measured data for the uniform mode (black dotted lines)[10, 30, 34].

Thus the spin dynamics of the nano-wire under the application of dc and rf currents can be understood through the Hall voltage rectification effect as discussed below.

### B. Additional Hall voltage induced by the application of the dc current

The inset of Fig. 3(a) shows the rf frequency dependence of the output signal $V_{\text{Hall}}$ for the dc currents from $I_{\text{dc}}$ = -12 (pink solid line) to +12 mA (black solid line) at every 6 mA in $H_{\text{ext}}$ = 500 Oe at $\phi = 30^{\circ}$. The sense of the dc current is defined as positive along the +*x* direction. For clarity, the non-resonant background signal larger exceeding the resonant signal is subtracted. As shown in the inset of Fig. 3(a), the spectrum for $I_{\text{dc}} = 0$ (red solid line) has at least two distinct modes near 6.8 and 7.5 GHz, while the spectrum for $I_{\text{dc}} = \pm 12$ mA has an additional distinct mode near 11.0 GHz. Figure 3(a) shows the variation of the rectified Hall signal:

$$\Delta V_{\text{Hall}}\left(\omega_{\text{k}}, I_{\text{dc}}\right) = V_{\text{Hall}}(I_{\text{dc}}) - V_{\text{Hall}}(I_{\text{dc}} = 0\,\text{mA}) = V_{\text{baseline}} + \Delta V_{1(2)}\,. \tag{46}$$

$V_{\text{baseline}}$ and $\Delta V_{1(2)}$ represent the baseline voltage and resonant Hall voltage difference, respectively. It is focused on the $I_{\text{dc}}$ dependence of the peak height at $\omega_k / 2\pi$ = 11 GHz. As shown in Fig. 3(b), the peak height is proportional to $I_{\text{dc}}$ as expected for a conventional resistive effect. In other words, the Hall voltage difference $\Delta V_2$ is expressed by $\Delta V_2 = \Delta R_{\text{PHE}} \cdot I_{\text{dc}}$, where $\Delta R_{\text{PHE}}$ is the planer Hall resistance. The estimated $\Delta R_{\text{PHE}}$ is 0.032 mΩ for $\phi = 45^{\circ}$ and $\left|\boldsymbol{H}_{\text{ext}}\right| = 0.5\,\text{kOe}$, and –0.053 mΩ for $\phi = 120^{\circ}$ and $\left|\boldsymbol{H}_{\text{ext}}\right| = 0.2\,\text{kOe}$. It

should be noted that both the sense and magnitude of $\Delta R_{\text{PHE}}$ strongly correlate with the direction of the magnetization and the precessional angle. The latter relationship between $\Delta R_{\text{PHE}}$ and the precessional angle is discussed in Section IV D.

### C. Excited spin-wave modes

The magnetic field dependence of the SW frequency for each spin mode is shown in Fig. 4. All observed modes are attributed to the magnetic excitations as discussed above. In particular, the two modes observed in the lower frequency region correspond to the quantized SW modes derived from the dipole-dipole interaction (red circles) and the dipole-exchange coupling (blue squares), due to the confined structure as discussed by Guslienko *et al.*[28] Another mode in the higher frequency region is the PSSW mode[1, 2, 35] (black triangles). An empirical expression describing the complete SW modes with the quantized integer numbers is evaluated below.

According to Guslienko *et al.*[28] and Bayer *et al.*[2], the frequency of the quantized SW modes (or Eigen-modes) of a strip can be evaluated from the solution of the LLG equation and is given by

$$\left(\frac{\omega_n}{\omega_M}\right)^2 = \left(\frac{\omega_H}{\omega_M} + 1 + \frac{\lambda_n}{4\pi}\right)\left(\frac{\omega_H}{\omega_M} + Aq_{\|}^2 - \frac{\lambda_n}{4\pi}\right), \tag{47}$$

where $A$ is the exchange stiffness coefficient, $q_{\|}$ is the in-plane wave vector, $\omega_H = \gamma_0 H_{\text{ext}}$,

$\omega_M = \gamma_0 4\pi M_S$ , and the dipole Eigen-value $\lambda_n$ is precisely given by Eq. (4) and approximately by the analytical expression of Eq. (12) as derived from Ref. [28].

In this case, an infinitely long magnetic strip is assumed, whose cross-section is a rectangular of thickness $d$ and width $w$ with $p = d/w \ll 1$. The SW is quantized along the width direction due to the dipole-dipole interaction under the pinning condition given by Eq. (8) in Ref. [28]. This provides $q_{\|} = n\pi / w_{\text{eff}}$ , where the integer $n$ is the quantization number and $w_{\text{eff}}$ is the effective wire width including the pinning effect at the edge of the strip.

The frequency variation in the PSSW mode with the applied field is qualitatively understood in terms of the rectangular cross-sectional strip. Accordingly to Kalinikos and Slavin[1, 2, 35], the dispersion relationship of the SW in a confined magnetic structure is given by:

$$\omega^2 = \gamma^2 \left( H + \frac{2A}{M_S} q^2 \right) \left( H + \frac{2A}{M_S} q^2 + 4\pi M_S \cdot F_{pp}\left( q_{\|} d_{\text{eff}} \right) \right), \tag{48}$$

where

$$q^2 = q_x^2 + q_y^2 + \left( \frac{m\pi}{d_{\text{eff}}} \right)^2 = q_{\|}^2 + q_{\perp}^2 \ , \tag{49}$$

and $d_{\text{eff}}$ is the effective thickness for the SW wavelength including the boundary conditions given by Eq. (8) in Ref. [28], $m$ the quantized number for the SW along the thickness direction and $F_{pp}\left( q_{\|} d \right)$ the matrix element of the magnetic dipole interaction[35].

The comparison between the experimental results and the analytical SW Eigen-frequencies estimated from Eqs. (47) and (48) for the two lowest modes ($n$ = 0 and 1) and another mode ($m$ = 1) is presented in Fig. 4. These results imply that the above analytical calculations correspond well to the present result. Consequently, the effective width $w_{\mathrm{eff}}$ and thickness $d_{\mathrm{eff}}$ are estimated at 5.2 μm and 65 nm respectively, both of which are almost equivalent to the measured width and thickness of the strip. This clearly indicates that the quantized SW is excited under the edge or surface pinning conditions. To simplify the rectification spectrum, a phenomenological macro-spin model as described by Eq. (9) is introduced, and the validity of the introduction of the uniform mode corresponding to the lowest quantized SW mode ($n$ = 0 and $m$ = 0) is examined. As shown in Fig. 4, the calculation of Eq. (32) qualitatively agrees with the present results of the field angle $\phi$ dependence of $\Delta V_{\mathrm{AMR}}$ and $\Delta V_{\mathrm{Hall}}$.

**D. Angle dependence of the additional Hall voltage due to the PSSW mode induced by the dc current**

As seen in Figs. 3(a) and 4, the spectra present at least three peaks centered around 6.2, 7.0 and 11.0 GHz. As discussed above, these frequencies are derived from the quantized SW. The rectangular Py strip essentially has the quantized SW modes in both lateral and horizontal directions. The two lower-frequencies SW modes correspond to the quantized SW modes in the lateral direction, and the highest-frequency SW mode is the PSSW mode in the horizontal direction respectively. In particular, the difference between the resonance frequency of the two former SW modes with the quantized indices $n$ = 0 and 1 is too small to be distinguished from the measurement of the angle dependence of the Hall voltage rectification spectra. In order to discuss the origin of the PSSW mode induced by the dc current and treat it as a form of macro-spin dynamics described by Eq. (9), the PSSW mode is focused upon, since the observed PSSW mode is only the lowest mode.

The dc voltage difference induced by the dc current, $\Delta V_{\mathrm{Hall}}$, given by Eq. (46) is plotted as a function of the applied magnetic field angle $\phi$ in Fig. 5. The subsequent relationship from Eqs. (42) and (46)[36], $\Delta V_{\mathrm{Hall}}$ is written as

$$\Delta V_{\mathrm{Hall}}\left(\omega_{\mathrm{k}}, I_{\mathrm{dc}}\right)=\frac{1}{2} I_{\mathrm{dc}} \Delta R\left[\sin 2\phi-2\left(1+\eta^{2}\right)\left(D_{0}^{2}+\frac{D_{1}^{2}+D_{2}^{2}}{2}\right)\sin 2\phi-2 D_{0}\sqrt{1+\eta^{2}}\cos 2\phi\right]. \quad (50)$$

The first term in Eq. (50) corresponds to the non-resonant baseline voltage $V_{\mathrm{base\,line}}=\left(I_{\mathrm{dc}}\Delta R\sin 2\phi\right)/2$. From the second and third terms, the Hall voltage difference and the planer Hall resistance are given by:

$$\Delta V_{2}=-I_{\mathrm{dc}}\Delta R\left[\left(1+\eta^{2}\right)\left(D_{0}^{2}+\frac{D_{1}^{2}+D_{2}^{2}}{2}\right)\sin 2\phi+D_{0}\sqrt{1+\eta^{2}}\cos 2\phi\right] \quad (51)$$

and

$$\Delta R_{\mathrm{PHE}}=\Delta R\left[\left(1+\eta^{2}\right)\left(D_{0}^{2}+\frac{D_{1}^{2}+D_{2}^{2}}{2}\right)\sin 2\phi+D_{0}\sqrt{1+\eta^{2}}\cos 2\phi\right]. \quad (52)$$

From Eqs. (36)-(38) and (51), the voltage amplitude $\Delta V_{2}$, induced by the magnetization dynamics in the PSSW mode, depends on either $\sin 2\phi$, $\sin 2\phi \cdot (\cos\phi)^2$, $\sin 2\phi \cos\phi$ or $\cos 2\phi$. In addition, the sense of the $\Delta V_2$ is opposite to that of $V_{\mathrm{baseline}}$. The applied field angle $\phi$ dependence of $\Delta V_2$ is influenced by the in-plane field, out-of-plane filed, adiabatic and non-adiabatic spin torques. The out-of-plane field provides a $\sin 2\phi$ contribution, while the in-plane driving field generates $\sin 2\phi \cdot (\cos\phi)^2$ or $\sin 2\phi \cos\phi$ contribution. The adiabatic and non-adiabatic spin torques along the major (*x*) axis produce $\sin 2\phi$ and $\cos 2\phi$ contributions, respectively. These predictions are confirmed by the measuring the angle $\phi$ dependence of $\Delta V_2$.

Figure 5(a) shows the variation of the non-resonant background baseline voltage $V_{\mathrm{baseline}}$ for $I_{\mathrm{dc}} = +15\,\mathrm{mA}$. The excellent agreement between the experimental data and the results of the $\sin 2\phi$ fitting curve confirms that the voltage $V_{\mathrm{baseline}}$ is derived from the time-independent first term of the right-hand side in Eq. (50) and that the dragging effects hardly affect the spin dynamics due to the non-collinear alignment of the magnetization[33], encouraging us to treat spin dynamics using the macro-spin model.

The experimental angle $\phi$ dependence of $\Delta V_2$ is shown in Figs. 5(b) and (c). As expected, the sense of $\Delta V_2$ is opposite to that of $V_{\mathrm{baseline}}$. The (black) dashed and (blue) dotted lines shown in Fig. 5(b) correspond to the $\sin 2\phi$ and $\sin 2\phi \cdot (\cos\phi)^2$ curves, respectively. The former is derived from the out-of-plane field contribution, while the latter corresponds to the in-plane field contribution. Conversely, the (black) dashed and (red) dotted lines shown in Fig. 5(c) are calculated based on the rf and dc spin torque including the adiabatic and non-adiabatic terms. As seen in Figs. 5(b) and (c), the present angle $\phi$ dependence of $\Delta V_2$ seems to be in agreement with the $\sin 2\phi$ fitting curve rather than the $\sin 2\phi \cdot (\cos\phi)^2$ and $\cos 2\phi$ curves. This present result demonstrates the positive agreement between the analytical calculation and the experiment. At this stage, however, it is difficult to distinguish experimentally between the rf fields and spin torques, since both the rf fields and spin torques include the same angle $\phi$ dependence of $\Delta V_2$ which is proportional to $\sin 2\phi$.

Further evaluation of each contribution is important to reveal the physical origin of the additional Hall voltage induced by the dc current. Initially, $\Delta R$ is estimated at $0.08\,\Omega$ from the result shown in Fig. 5(a) by using the relation $V_{\text{base line}} = \frac{1}{2} I_{\text{dc}} \Delta R \sin 2\phi$. By using the parameters, $\phi = 45^\circ$ and $\Delta R_{\text{PHE}} = 0.032\,\text{m}\Omega$ in Fig. 3(b), the amplitude $D_0^2 + \frac{D_1^2 + D_2^2}{2}$ is determined to be $2 \times 10^{-2}$.

The static magnetic fields due to the dc current $I_{\text{dc}} = 15\,\text{mA}$ flowing through the wire and the CPW electrodes are estimated as $h_{\text{in-plane}}^{\text{static}} = \frac{I}{2w} \approx 18.8\,\text{Oe}$ and $h_{\text{out-of-plane}}^{\text{static}} = \frac{I}{2y} \approx 1.88\,\text{Oe}$ [37], respectively. The static field induced by the dc current, $h_{\text{in-plane}}^{\text{static}}$, has additional significant effects on the field distribution and the magnetization dynamics. Consequently, the dynamic field distribution of the in-plane and out-of-plane dynamic demagnetizing-field is distorted by the dc current, the magnitude of which is assumed to be $h_{\text{in-plane}} \approx h_{\text{out-of-plane}} \approx 1\,\text{Oe}$. On the other hand, $\partial\phi/\partial x$ from the precessional angle of the uniform mode in the 10 μm long ferromagnetic wire is evaluated to be $\frac{5^\circ}{10\,\mu\text{m}} = 0.87\times10^4\ \text{rad/m}$ [38]. On the other hand, the derivative $\partial\phi/\partial y$ dominates in the higher order SW excitation state described by Eq. (47), where the two-dimensional spin variation contains a large DW or higher order SW excitation. Therefore, the term $\partial\phi/\partial x$ is estimated for the following case: $\gamma_0 = 2.8\text{MHz/Oe}$, $\Delta = 30\,\text{GHz}$, $\omega_\text{k} = 6.3\,\text{GHz}$, $\gamma_0 H'_\text{c} = 29\,\text{GHz}$, $\alpha = 0.01$, $u_\text{dc} = 2.26\,\text{m/s}$ and $u_\text{rf} = 0.7\,\text{m/s}$, corresponding to $I_\text{dc} = 15\,\text{mA}$ and $I_\text{rf} = 4.8\,\text{mA}$ by assuming $\beta = 0$ and $P = 0.7$ [27]. Accordingly, the adiabatic spin-transfer torques are given by $u_\text{dc}\frac{1}{\alpha\Delta}\cdot\frac{\partial\phi}{\partial x} = 6.5\times10^{-5}$ and $u_\text{rf}\frac{1}{\alpha\Delta}\cdot\frac{\partial\phi}{\partial x} = 2.0\times10^{-5}$. The out-of-plane field is $\frac{\gamma_0 h_{\text{out-of-plane}}}{\alpha\Delta} = 9.3\times10^{-3}$, the in-plane fields $\frac{\alpha\gamma_0 h_{\text{in-plane}}}{\alpha\Delta} = 9.3\times10^{-5}$ and $\frac{\gamma_0 H'_\text{c}\gamma_0 h_{\text{in-plane}}}{\omega_k\alpha\Delta} = 4.3\times10^{-2}$. It is appropriate to compare each contribution with the amplitude $D_0^2 + \frac{D_1^2 + D_2^2}{2}$. As a result, that of the spin-transfer torque is smaller than that of the rf field even when $h_{\text{in-plane}} \approx h_{\text{out-of-plane}} \approx 1\,\text{Oe}$. Therefore, the additional Hall voltage induced by the dc current is ascribed to the dynamic field distorted by the dc current.

The contributions of driving torques have now been evaluated. The contribution of the rf field is dominant in the Py strip without DWs. The present analytical calculation is useful to evaluate the contributions of spin torques in ferromagnetic conductors with DWs. In a twisted spin structure of $180^\circ$ DW 300 nm in width, $\frac{\partial \phi}{\partial x}$ is about $\frac{180^\circ}{300\,\mathrm{nm}} = 1.05\times 10^{7}\ \mathrm{rad/m}$. The contribution of the adiabatic spin-transfer torque is $u_{\mathrm{dc}}\frac{1}{\alpha\Delta}\cdot\frac{\partial \phi}{\partial x} = 7.9\times 10^{-2}$ and $u_{\mathrm{rf}}\frac{1}{\alpha\Delta}\cdot\frac{\partial \phi}{\partial x} = 2.5\times 10^{-2}$, indicating that the spin-transfer torque controls the magnetization dynamics. This is also applicable to the $\partial\phi/\partial y$ for the higher order SW excitation, which is hard to describe using the one-dimensional spin configuration model.

The present Hall voltage method offers a highly sensitive detection of the spin dynamics induced by the rf current and that the existence of the mixing term $\delta V_{\mathrm{dc\cdot rf}}^{\mathrm{Hall}}$ is firstly observed. This represents the fact that *the individual contributions of the driving torque are distinguished by using the present method and that it* opens a new path to understanding the instabilities of ferromagnetism under spin-current flow [21, 26, 27].

## V. CONCLUSION

Highly sensitive measurements on dc planer Hall voltage spectra in a micron-scale single-layered $Ni_{81}Fe_{19}$ strip were performed as functions of frequency, external static field, field direction and dc current. Dynamic change in the rectification spectrum by the dc current was found. Based on a phenomenological analytical model, the changes produced by the inhomogeneous magnetic distribution under the coexistence of the dc current and the spin-torque were quantitatively evaluated. This highly sensitive detection of small spin dynamics via the planer Hall rectification effect represents a powerful technique for studying spin dynamics within a single nano- or micron-scale confined magnetic structure, and provides a way to understand the detailed correlation between a localized magnetic moment and a conduction electron.

## VI. ACKNOWLEDGEMENTS

The present study was partly supported by MEXT Grants-in-Aid for Scientific Research in a Priority Area, the JST PREST, and a JSPS Grants-in-Aid for Scientific Research. A. Y. also acknowledges support from the JST PRESTO program.

## APPENDIX: DETAILED SOLUTION OF THE LLG EQUATION

In the approach, magnetization dynamics are solved by Eq. (12). The details of each term in Eq. (12) are defined as follows:

$$\frac{d\boldsymbol{m}}{dt}+\boldsymbol{\Omega}\times\boldsymbol{m}=e^{i\omega t}\begin{pmatrix}-\Omega_0 m_b\\ i\omega m_b+\Omega_0 e^{-i\omega t}\\ i\omega m_c\end{pmatrix}, \tag{A1}$$

$$\begin{aligned}-\gamma_0\boldsymbol{m}\times\left(\boldsymbol{H}_{\mathrm{ext}}+\boldsymbol{H}_{\mathrm{A}}\right)&\approx-\gamma_0\begin{pmatrix}-m_c e^{i\omega t}H_{\mathrm{ext}}\sin\left(\phi_0-\phi\right)\\ m_c e^{i\omega t}H_{\mathrm{ext}}\cos\left(\phi_0-\phi\right)\\ H_{\mathrm{ext}}\sin\left(\phi_0-\phi\right)-m_b e^{i\omega t}H_{\mathrm{ext}}\cos\left(\phi_0-\phi\right)\end{pmatrix}\\ &-\gamma_0\begin{pmatrix}m_c e^{i\omega t}\frac{1}{2}M_S\left(N_x-N_y\right)\sin 2\phi\\ m_c e^{i\omega t}M_S\left[N_z-\left(N_x\cos^2\phi+N_y\sin^2\phi\right)\right]\\ -\frac{1}{2}M_S\left(N_x-N_y\right)\sin 2\phi-m_b e^{i\omega t}\left(N_y-N_x\right)\cos 2\phi\end{pmatrix}, \\ &=-\gamma_0\begin{pmatrix}-m_c e^{i\omega t}H'_{\mathrm{eff}}\\ m_c e^{i\omega t}H'_c\\ H'_{\mathrm{eff}}-m_b e^{i\omega t}H'_b\end{pmatrix}\end{aligned} \tag{A2}$$

$$-\gamma_0\boldsymbol{m}\times\boldsymbol{h}_{\mathrm{rf}}=e^{i\omega t}\begin{pmatrix}0\\ \gamma_0 h_{\mathrm{out}}\\ -\gamma_0 h_{\mathrm{in}}\cos\phi\end{pmatrix}, \tag{A3}$$

$$\alpha\boldsymbol{m}\times\left(\frac{d\boldsymbol{m}}{dt}+\boldsymbol{\Omega}\times\boldsymbol{m}\right)=e^{i\omega t}\begin{pmatrix}-\alpha\Omega_0 m_c\\ -i\omega\alpha m_c\\ i\omega\alpha m_b+\Omega_0\alpha e^{-i\omega t}\end{pmatrix}, \tag{A4}$$

In the $(x,y,z)$ coordinate system, the spin-polarized current directed along the $x$-axis is given by $\boldsymbol{u}=\frac{P\mu_B}{eM_S}j\left(1,0,0\right)$. The unit vector along the effective field is described by $\boldsymbol{m}=\left(\sin\theta\cos\phi,\sin\theta\sin\phi,\cos\theta\right)$. Subsequently, the adiabatic and non-adiabatic spin-transfer torque terms are respectively derived as

$$-(\boldsymbol{u}\cdot\nabla)\boldsymbol{m} = -\frac{P\mu_{\mathrm{B}}}{eM_{\mathrm{S}}} j\cdot\frac{\partial}{\partial x}\begin{pmatrix}\cos\phi\\ \sin\phi\\ 0\end{pmatrix} = -\frac{P\mu_{\mathrm{B}}}{eM_{\mathrm{S}}} j\cdot\begin{pmatrix}-\sin\phi\cdot\frac{\partial\phi}{\partial x}\\ \cos\phi\cdot\frac{\partial\phi}{\partial x}\\ 0\end{pmatrix} \tag{A5}$$

and

$$\beta\boldsymbol{m}\times\left[(\boldsymbol{u}\cdot\nabla)\boldsymbol{m}\right] = \frac{\beta P\mu_{\mathrm{B}}}{eM_{\mathrm{S}}} j\begin{pmatrix}0\\ 0\\ \frac{\partial\phi}{\partial x}\end{pmatrix}. \tag{A6}$$

Both spin torque terms in the $(a,b,c)$ coordinate system can be obtained by a rotation transformation from the $(x,y,z)$ coordinate system to the $(a,b,c)$ coordinate system;

$$-(\boldsymbol{u}\cdot\nabla)\boldsymbol{m} = -\frac{P\mu_{\mathrm{B}}}{eM_{\mathrm{S}}} j\begin{pmatrix}0\\ \frac{\partial\phi}{\partial x}\\ 0\end{pmatrix} \tag{A7}$$

and

$$\beta\boldsymbol{m}\times\left[(\boldsymbol{u}\cdot\nabla)\boldsymbol{m}\right] = \beta\frac{P\mu_{\mathrm{B}}}{eM_{\mathrm{S}}} j\begin{pmatrix}0\\ 0\\ \frac{\partial\phi}{\partial x}\end{pmatrix}. \tag{A8}$$

**[References]**

**[Figure captions]**

**Figure 1**

(a) Schematic diagram of the rf measurement, including an overhead view of the optical micrograph of the device, and (b) the corresponding model geometry and symbol definitions. (c) Schematic projection of the magnetic moment precession in (*a*, *b*, *c*) coordinate axes.

**Figure 2**

(a) Typical rectifying AMR and Hall spectra in the absence of the dc current measured under the external static field of 500 Oe at $\phi = 45^{\circ}$. The amplitudes of both $\Delta V_{\mathrm{AMR}}$ and $\Delta V_{\mathrm{Hall}}$ are defined as the voltage difference between the peak and dip in the FMR frequency. (b) $\Delta V_{\mathrm{AMR}}$ and (c) $\Delta V_{\mathrm{Hall}}$ as a function of the field angle $\phi$ under $H_{\mathrm{ext}} =$ 100 [(blue) squares] and 500 Oe [(red) circles]. Fitting lines with Eqs. (44) and (45) are also shown.

**Figure 3**

(a) Typical Hall voltage difference $\Delta V_{\mathrm{Hall}}$ spectra, given by $\Delta V_{\mathrm{Hall}} = V_{\mathrm{Hall}}(I_{\mathrm{dc}}) - V_{\mathrm{Hall}}(0\,\mathrm{mA})$, measured under $H_{\mathrm{ext}} = 500$ Oe at $\phi = 30^{\circ}$. The inset shows the entire Hall voltage $V_{\mathrm{Hall}}$ spectra. (b) The dc current dependence of the Hall voltage difference $\Delta V_2$.

**Figure 4**

Magnetic field dependence of the resonant frequency observed by the Hall voltage spectra. The lines show the calculations as described in Ref. [28] for the quantized SW mode with quantized indices $n$ = 0 (black dashed line) and 1 (black dotted line), as well as the uniform mode [(red) dashed line] and the PSSW mode with the lowest quantized number (black solid line labeled PSSW). (Red) circles, (blue) squares and black triangles represent the experimental results of three peaks in the Hall voltage spectrum as typically shown in the inset.

**Figure 5**

Magnetic field angle $\phi$ dependence of (a) the (non-resonant) baseline Hall voltage $V_{\mathrm{base\,line}}$ and (b), (c) the Hall voltage difference $\Delta V_2$ induced by $I_{\mathrm{dc}} = +15\,\mathrm{mA}$ at the PSSW frequency under $H_{\mathrm{ext}} = 500\,\mathrm{Oe}$. The (black) dashed, (blue) dotted and (red) broken lines correspond to the fitting $\sin 2\phi$, $\sin 2\phi(\cos\phi)^2$ and $\cos 2\phi$ lines derived from each contribution, respectively.

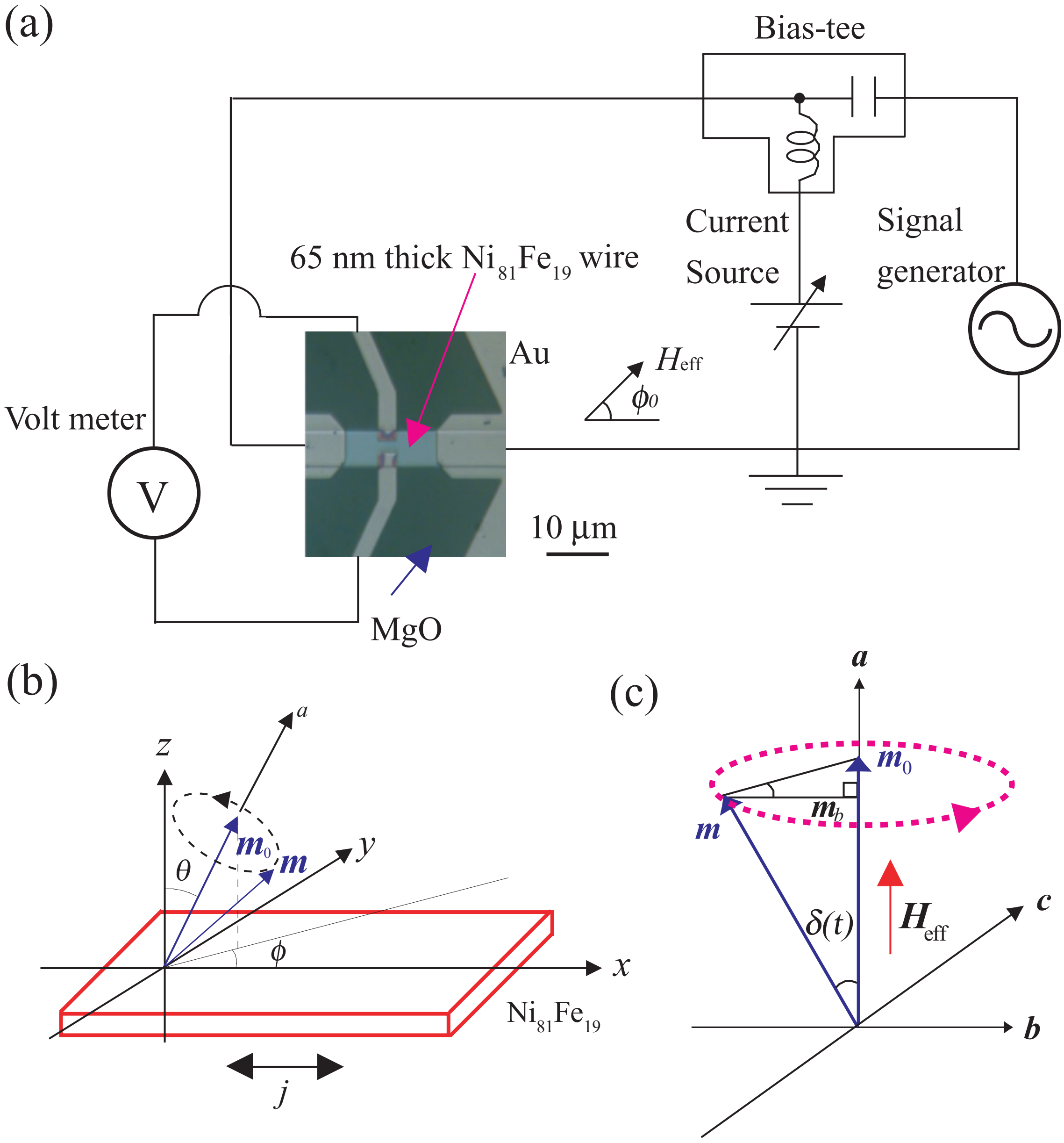

Fig. 1

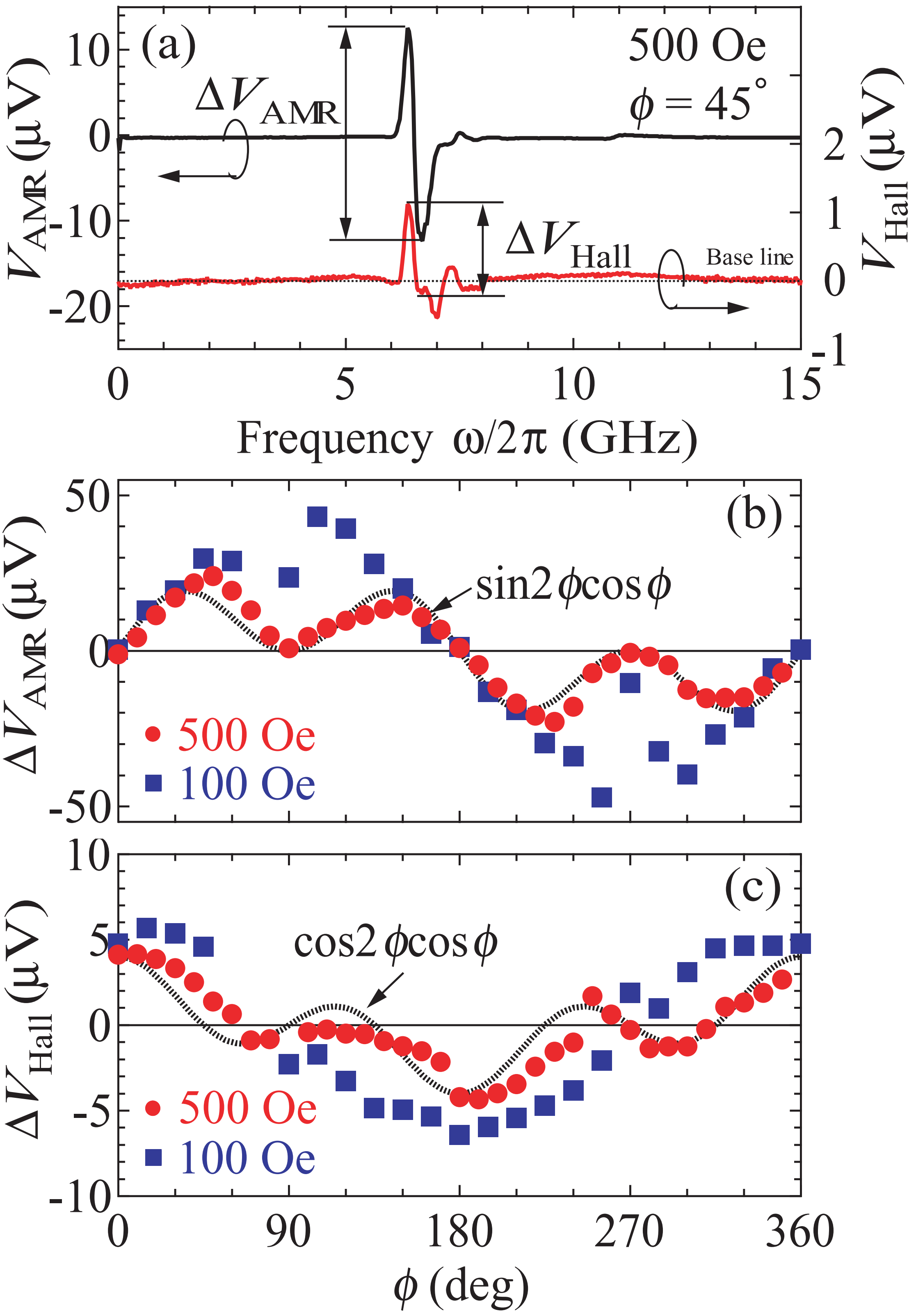

Fig. 2

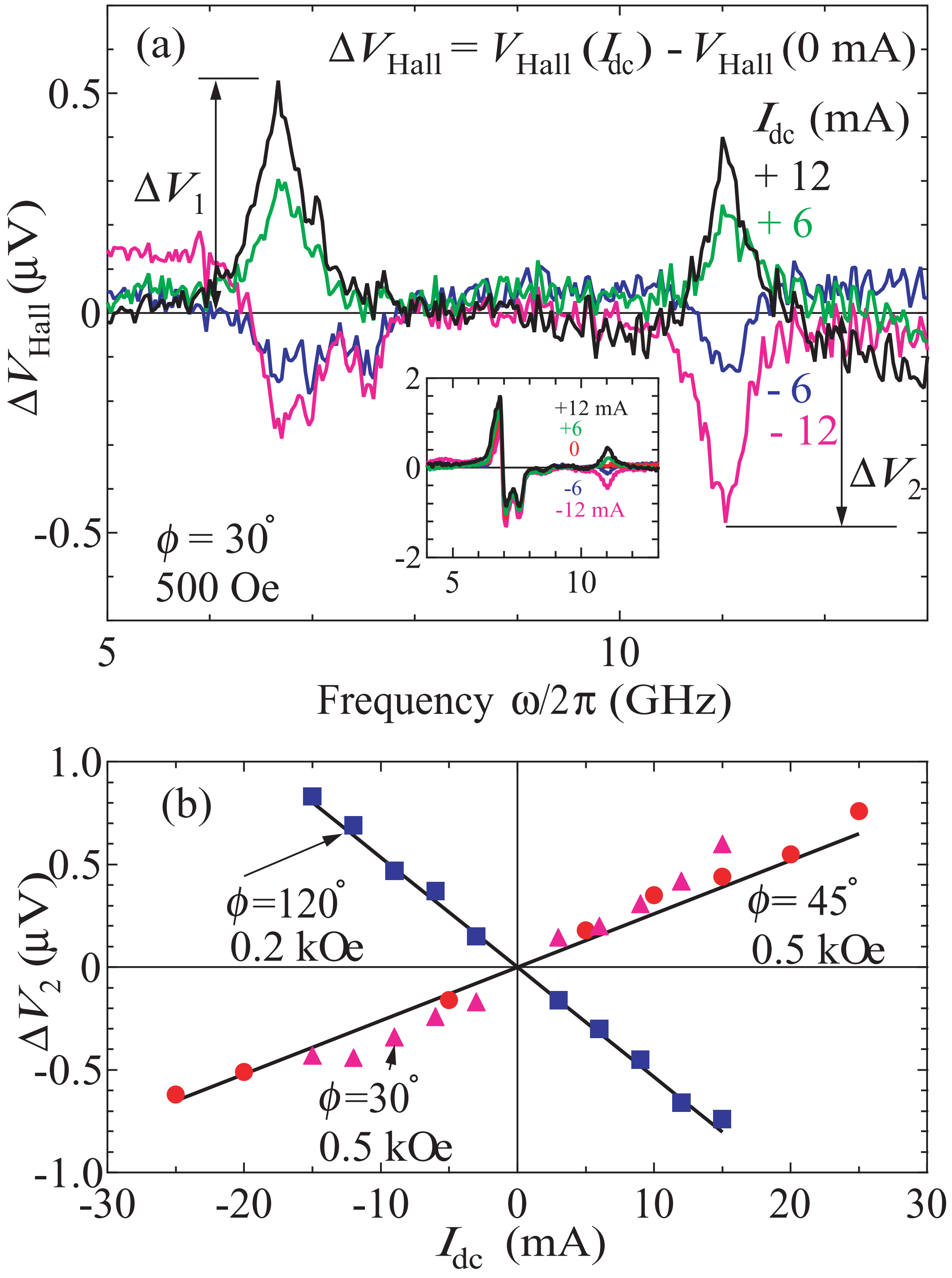
(a)
$\Delta V_{\mathrm{Hall}} = V_{\mathrm{Hall}}(I_{\mathrm{dc}})$ - $V_{\mathrm{Hall}}$ (0 mA)
$I_{\mathrm{dc}}$ (mA)
+ 12
+ 6
- 6
- 12
$\Delta V_1$
$\Delta V_2$
$\phi = 30^{\circ}$
500 Oe
+12 mA
+6
0
-6
-12 mA
$\Delta V_{\mathrm{Hall}}$ (μV)
Frequency ω/2π (GHz)
(b)
$\phi$=120°
0.2 kOe
$\phi$= 45°
0.5 kOe
$\phi$=30°
0.5 kOe
$\Delta V_2$ (μV)
$I_{\mathrm{dc}}$ (mA)

Fig. 3

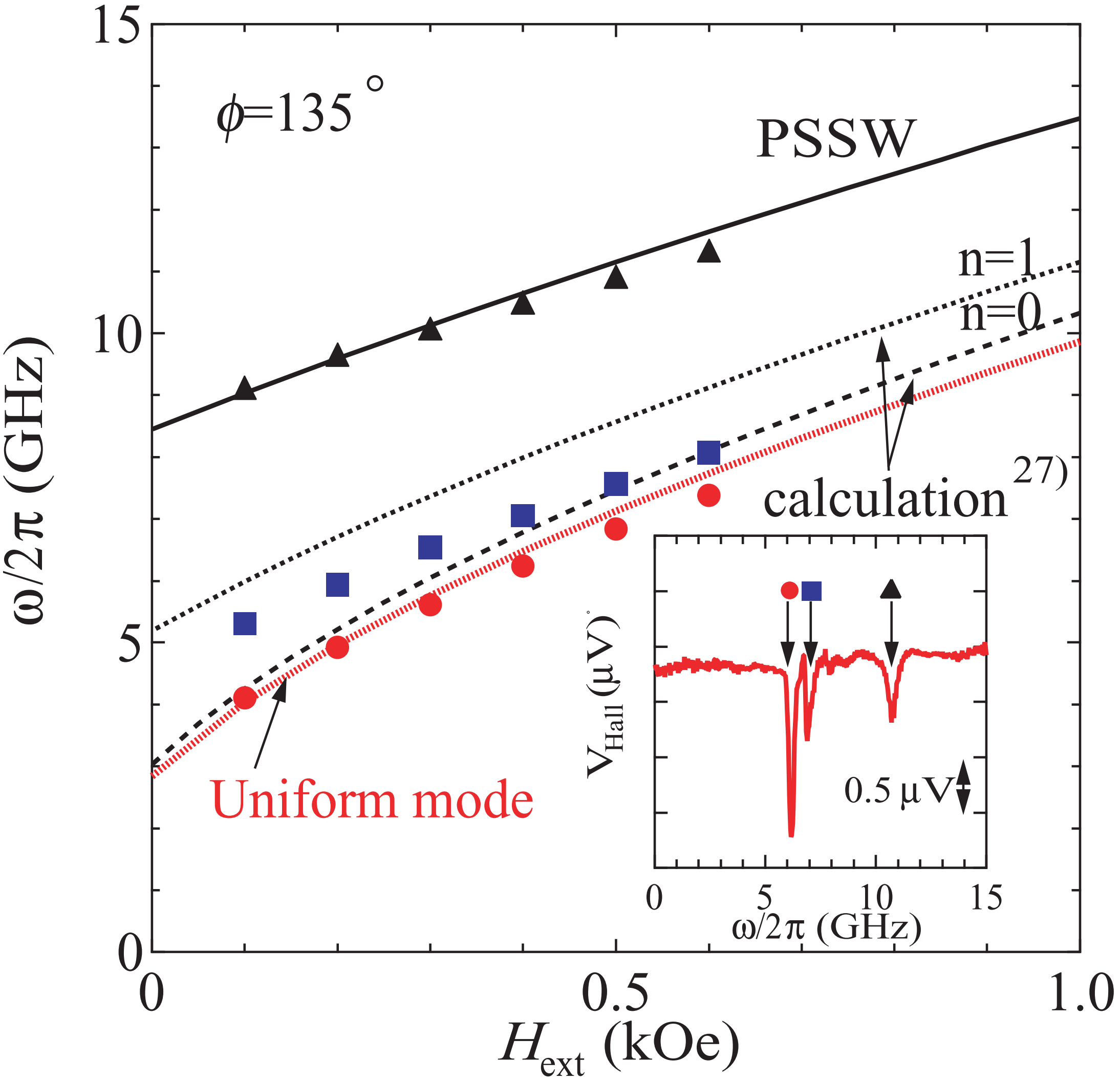

Fig. 4

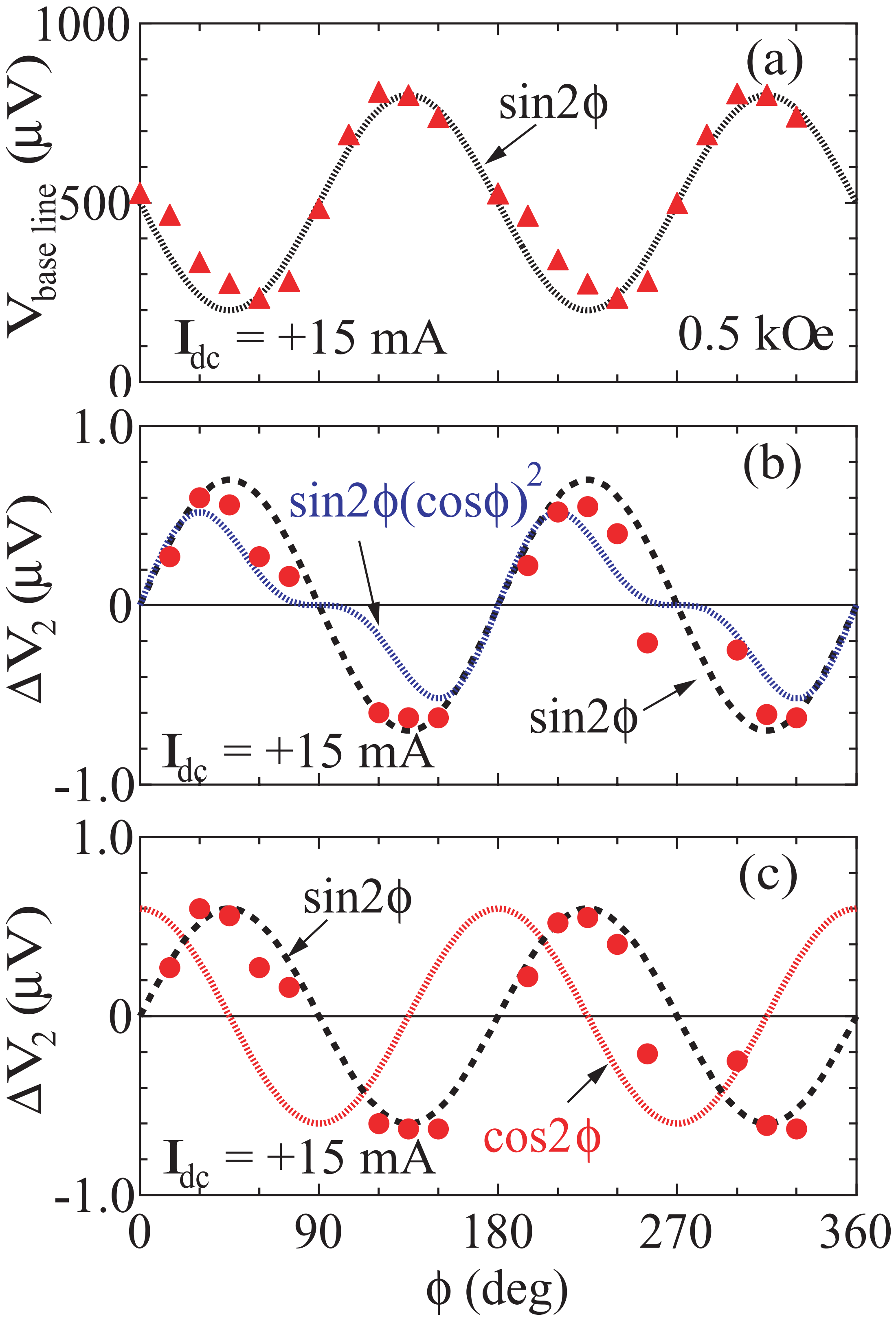

Fig. 5